# Designing a Framework to Develop WEB Graphical Interfaces for ORACLE Databases - Web Dialog

**Georgiana-Petruța Fîntîneanu**
**Florentina Anica Pintea,**
**Faculty of Computers and Applied Computer Science,**
**"Tibiscus" University of Timişoara**

**ABSTRACT**. The present article aims to describe a project consisting in designing a framework of applications used to create graphical interfaces with an Oracle distributed database. The development of the project supposed the use of the latest technologies: database Oracle server, Tomcat web server, JDBC (Java library used for accessing a database), JSP and Tag Library (for the development of graphical interfaces).

**Project Presentation**

Basically, the project offers a library of user friendly tags for a person with limited knowledge of programming, implementing a number of features needed to develop a graphical interface with a database.

The development of a form that has access to a database requires the creation of text files that resembles a lot the files in HTML format. The interfaces that can be developed allow the bidirectional communication with the database (display data, call stored procedures) and a transmission of values between the objects from the forms. The main advantage of using this project is the possibility of providing the secure, quick and clean access to a database distributed via Internet.

This paper contains a brief presentation of the technologies used, focusing on presenting the tags library. It is presented the architecture of the system and the communication between the component parts of the system.

The general scheme of the application is the following:

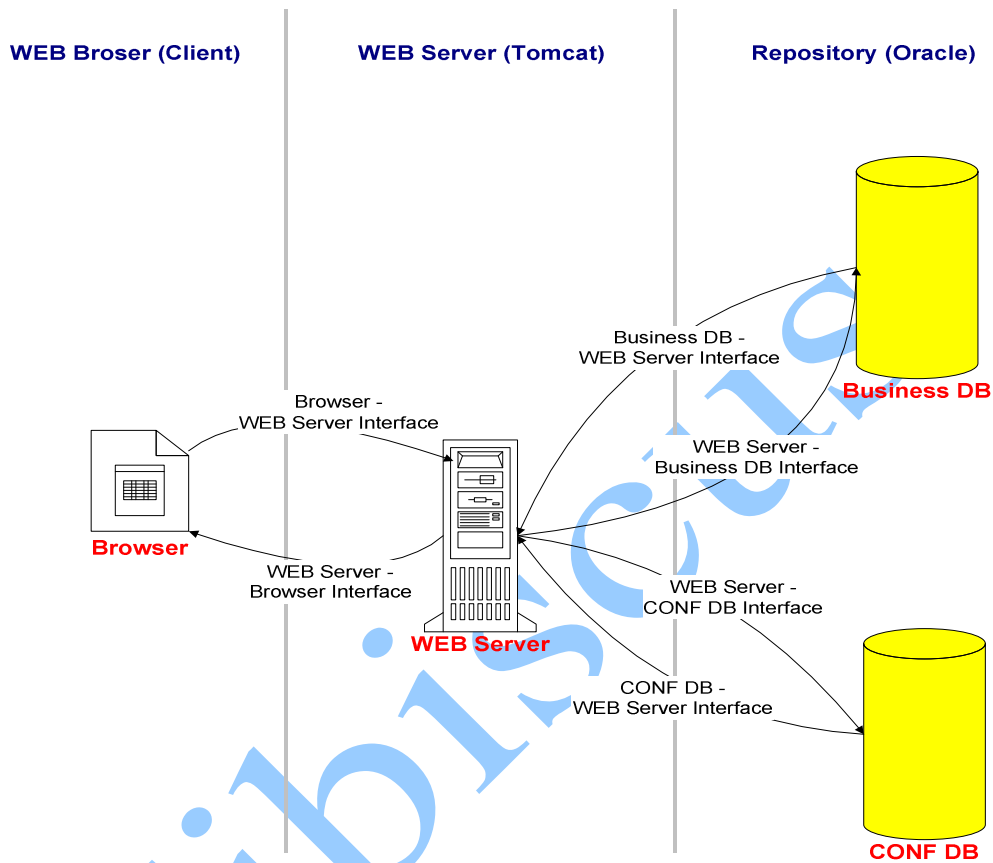

Figure 1

As it can be seen from the figure the application has three components:
1. The storage part which is the Oracle server database where the data is permanently saved.
2. The WEB server is the intermediate component that solves all requests and responses between the other components of the system. The WEB server is also used as an internal "cache" for the

      static data (configurations, access rights) to avoid the querying of the database when it is not necessary.
3. The third component is the WEB browser, where all requests are made by the user. The requests are made by the user through a visual element called dialog.

The programs package of the WEB dialog system must be installed and configured on each of the three components (especially on the Web server and database) to ensure the transfer of safe, easy and clean data. An advantage of the system lies in the fact that on the customer's part it is enough to install Internet Explorer 5.5 or a more recent version, produced by Microsoft.

The Products developed using the WEB Dialog system are named dialogs and consist of JSP pages. A JSP file contains HTML and XML codes as well as elements specific to the technology. When a JSP page from the WEB server is loaded, it is sent to a JSP "engine" which compiles the file generating a servlet that it is running. Following the running a HTML code is generated that is sent to the client browser. After the first loading of a JSP page, if the source file is not changed, the WEB server will not recompile it and will run the servlet that has already been generated. In addition, the system offers the possibility of compiling all the JSP files at the running of the application. It is very easy to develop a new dialog with the database due to the developed tag library. The main purpose of the tag library is to allow the author of the JSP page who has basic knowledge of HTML to build a page that offers a number of features by adding tags in JSP. For the author of the JSP page the Java code behind a tag is transparent as well as the fact that some tags run queries on the database.

The Java code necessary to implement the tag is simple, easily to be understood and maintained. Also a tag library makes the JSP programmer to split his work in well-defined components.

Behind each tag is a Java class that implements the functionality. For each tag we can define a set of attributes that represent the static properties. Through the Java class these attributes can be overwritten with values from the database.

An advantage of the tags is the possibility to specify the parent-child relationship, leading to the mandatory compliance of the tags order to load a JSP page.

The advantages of using a tag library can be summarized as follows:
- Simplicity;
- Modularized code;
- Small compilation steps;
- Easily documented.

The configuration of a dialog is saved on the Web server. The objects described by tags have a range of attributes that describes the properties of objects. The default values of properties are saved in the JSP files. The respective values may be modified depending on the user who is connected to the application or depending on the language with which the user is connected.

The part of the data storage is divided into two major components. These components are grouped according to the functionality they have in the WEB dialog system.

Business DB = Business Logical Database is the database for which the graphical interface is developed. Via the Web dialog system different types of data may become interchangeable between the user and the database.

CONF DB = Configuration Database (the database for configuration) is the repository of dynamic data from a dialog: the text saved in multiple languages, personalized features of some objects, access rights to objects.

The values of the features dependent on the user or on the language are kept on CONF DB. The architecture of the database configuration looks like:

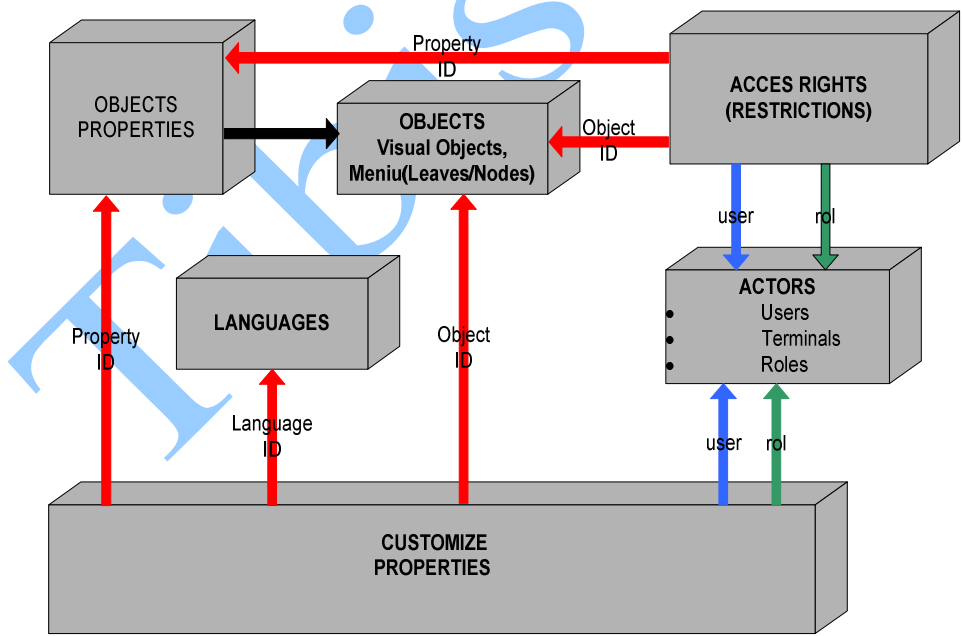

Figure 2

On CONF FB it is stored the information on all the users who have access to the database and the internet addresses of the terminals from where the connection is made.

For a dialog the identifiers of objects, the objects type and their properties are stored. By the feature of an object we mean an attribute of the tag which implements the object. Only the modified features for a particular user are saved in the database (the default values are saved in JSP files). If more users may have the same features for an object, different from the default ones, then the users can be grouped into groups of users called roles. The roles are associated with a user on priority levels. For a user with multiple associated roles, if the feature of an object has different values for different roles, it will be considered a saved valued for the role with the highest priority. Each user is automatically assigned to a role with the same identifier that will always be of top priority. Therefore the features saved at the user's level overwrite the features saved at the role level. Some features (all texts displayed) are dependent on the language to which the user is connected, so they may have different values for different languages and they are stored properly.

The saving of a dialog in the database is made from a client computer by a user who has the administrator right. Any change of the personalized features can be done by any users.

CONF DB is also used to store the history of the performed actions and information on alert signals sent to the client.

**Conclusions**

The advantages of Web dialog:
- ➢ Easily to use - people who develop dialogs do not need specialized programming knowledge;
- ➢ Storage configuration dialogs is made in one place only (Web server);
- ➢ Opening a dialog does not query the database only if it is absolutely necessary;
- ➢ It allows the encapsulation of the objects;
- ➢ It requires minimal installation at the client's level (browser);
- ➢ Slight possibility of implementing new requirements (Upgrade of the system) and the implementation of special requirements;
- ➢ The access to Business DB can be done directly;
- ➢ The access to the applications can be made via the Internet.

Disadvantages:
- ➢ The technology requires hardware resources which are rather expensive;
- ➢ It is possible to limit a maximum number of objects on the dialog so that the generated java class for JSP file does not exceed 64k. The limitation is due to the Java virtual machine and it depends on its implementation and on the operating system that is installed.